\documentclass[12pt]{article}
\usepackage{amsmath,txfonts,amssymb,bm,array,graphics}
\newcommand{\bequ}{\begin{equation}}
\newcommand{\nequ}{\end{equation}}
\begin{document}
\sloppy
\input epsf
\begin{center}
\large{\bfseries {Viscosity of Earth's Outer Core}}\normalsize$^\dagger$ \\

\bigskip

D. E. SMYLIE$^{1}$ and Andrew Palmer$^{2}$ \\
\bigskip

\noindent
$^{1}$Department of Earth and Space Science and Engineering, York University \\
4700 Keele Street, Toronto, Ontario, M3J 1P3, CANADA \\
Phone:(416) 736-2100, ext. 66438, Fax:(416) 736-5817\\
E-mail: doug@core.yorku.ca \\
\bigskip

\noindent
$^{2}$Department of Physics and Astronomy, York University \\
4700 Keele Street, Toronto, Ontario, M3J 1P3, CANADA \\
E-mail: palmer@core.yorku.ca \\
\bigskip

\end{center}
\bigskip
\footnotetext{$^{\dagger}$Published electronically in arXiv.org$>$physics
$>$physics.geo-ph, Cornell University Library, Ithaca, N. Y., \today.}

\begin{abstract}
\noindent
A viscosity profile across the entire fluid outer core is found by
interpolating between measured boundary values, using a differential
form of the Arrhenius law governing pressure and temperature dependence.
The discovery that both the retrograde and prograde free core nutations are
in free decay (Palmer and Smylie, 2005) allows direct measures of viscosity
at the top of the outer core, while the reduction in the rotational splitting
of the two equatorial translational modes of the inner core allows it to be
measured at the bottom. We find $2,371\pm 1,530\;Pa\cdot s$ at the top and
$1.247\pm 0.035\times 10^{11}\;Pa\cdot s$ at the bottom.\\

\noindent
Following Brazhkin (1998) and Brazhkin and Lyapin (2000) who get $10^2\;Pa\cdot
s$ at the top, $10^{11}\;Pa\cdot s$ at the bottom, by an Arrhenius
extrapolation of laboratory experiments, we use a differential form of the
Arrhenius law to interpolate along the melting temperature curve to find a
viscosity profile across the outer core. We find the variation to be closely
log-linear between the measured boundary values.\\

\noindent
The close agreement of the boundary values of viscosity, found by
Arrhenius extrapolation of laboratory experiments, with those found from the
free core nutations, and the inner core translational modes, suggests that core
flows are laminar and that the returned viscosities are measures of their
molecular values. This would not be the case in the presence of the vigourous
turbulent convection sometimes postulated by dynamo theorists.\\

\noindent
The local Ekman number is found to range from $10^{-2}$ at the bottom of
the outer core to $10^{-10}$ at the top. Except in the very lower part of the
outer core, Ekman numbers are in the range $10^{-4}$ to $10^{-5}$, or below, in
which the laminar flows of numerical dynamos and laboratory rotating fluids
experiments occur.\\   
  
\noindent
We find explicit expressions for the reciprocal $Q$'s at both boundaries and
for the viscous coupling torques between the outer core and shell, and between
the outer and inner cores. For the high viscosity in the F-layer outside the
ICB found from the reduction in the rotational splitting of the two
equatorial translational modes of the inner core, the inner core is found to be
tightly coupled to the outer core with a negligible contribution to dissipation
in the free core nutation modes.

\end{abstract}

\section{Introduction}

   Properties of Earth's deep interior, such as its elasticity, density,
pressure and gravity have traditionally been obtained through the inversion
of seismic observations. While these have been important to our understanding
of Earth's internal structure, the viscosity of the outer fluid core is
crucial to our understanding of its dynamics and the generation of the
geomagnetic field. Direct observations and limits on viscosity have
traditionally been much larger than those values found on the basis of
extrapolations of laboratory high pressure and temperature experiments
(Lumb and Aldridge, 1991). The latter tend to be close to that of liquid iron
at atmospheric pressure (Rutter et al., 2002), while the former are many orders
of magnitude larger (Davis and Whaler, 1997).

   Unusual properties of the lower outer core have long been suspected, going
back to the 1926 claim by Jeffreys (Jeffreys, 1926) of a strong negative
P-wave velocity gradient there. While the P gradient is now thought to be
small, but slightly positive, possibly due to solid inclusions slowing
compressional waves in that region (Garland, 1971, pp.42-50), its properties
remain a subject of speculation. A new seismic phase, PKhKP, has even been
suggested by Bolt (Bullen and Bolt, 1985, p.317) as originating from
reflections in the lower outer core. 

   With the discovery that both the retrograde and prograde free core nutations
are in free decay (Palmer and Smylie, 2005), direct measures of the viscosity
at the top of the outer core can be made. In this paper, we present a detailed
analysis of both the Ekman layer at the top of the outer core and that at the
bottom, just outside the inner core boundary. Our analysis yields a mean value
of $2,371\pm 1,530\;Pa\cdot s$ for the recovered dynamic viscosity at the top
of the outer core. From the reduction in the
rotational splitting of the two equatorial translational modes of the inner
core, two independent measures of viscosity in the F-layer at the bottom of
the outer core are found (Smylie, 1999; Smylie and McMillan, 2000), as the
reduction is larger for the retrograde mode than for the prograde mode. The
retrograde equatorial mode gives $1.190\pm 0.035\times 10^{11}\;Pa\cdot s$,
while the prograde equatorial mode gives $1.304\pm0.034\times 10^{11}\;Pa\cdot
s$ for an average value of $1.247\pm 0.035\times 10^{11}\;Pa\cdot s$.

   Our values of viscosity at the boundaries of the outer core are in close
agreement with an Arrhenius extrapolation of laboratory experiments by
Brazhkin (1998) and by Brazhkin and Lyapin (2000) who find $10^2\;Pa\cdot s$
at the top of the outer core and $10^{11}\;Pa\cdot s$ at the bottom. Although
the Arrhenius model is widely used over the limited ranges of pressure and
temperature involved in laboratory experiments (Dobson, 2002), its exponential
nature yields values of viscosity that appear extreme when extrapolated from
laboratory pressures of the order of $30\;GPa$ by more than a factor of ten
to outer core pressures, and when the temperature extrapolation from laboratory
values of the order of $2,000\;K$ is by a factor of two to outer core melting
temperatures. Our measured viscosities at the boundaries appear to confirm the
validity of Arrhenius extrapolation. 

   The close agreement of the boundary values of viscosity with an Arrhenius
extrapolation of laboratory experiments prompts us to construct a viscosity
profile across the entire outer core using the Arrhenius law. Due to the
strong pressure dependence of the activation volume, it is necessary to use
a differential form of the Arrhenius law to interpolate along the melting
temperature curve between the measured boundary values of viscosity.

   Both the free core nutations and the translational oscillations of the
inner core are highly influenced by Earth's rotation. They are examples of
dynamical phenomena in contained rotating fluids. This is a subject that was
of much interest several decades ago based on the inertial wave equation
(Stewartson and Roberts, 1963; Roberts and Stewartson, 1965; Greenspan, 1969;
Busse, 1968; Aldridge and Toomre, 1969). In these descriptions, the container
rotation is constrained rather than free, there is no inner body, the fluid is
assumed incompressible, uniform and non self-gravitating and the container is
taken to be rigid. Although the theories were beautifully confirmed by
laboratory experiments (Aldridge, 1967) their relevance to the real Earth in
which the rotation of the container is unconstrained, there is an inner body,
the fluid is compressible, stratified and self-gravitating and the boundaries
are deformable does not appear to be direct. The governing equation then
becomes the subseismic wave equation rather than the inertial wave equation
(Smylie, Jiang, Brennan and Sato, 1992). The free core nutations are solutions
of the subseismic wave equation which are close to pure rotations with respect
to the shell, though not exactly so (Jiang, 1993; Jiang and Smylie, 1995;
Jiang and Smylie, 1996). Thus, the motions in the Earth frame
imitate the `spin-over' mode of the inertial wave equation. Some of the basic
features of the Ekman layers at the two boundaries in the real Earth resemble
those of the `spin-over' mode but previous theories are not given in detail
with only results shown, possibly due to Greenspan's dictum that the
`computation of viscous effects, ..., is laborious but straightforward'
(Greenspan, 1969, p.66). We give full details of the analysis of the Ekman
layers at the two boundaries independent of the inertial wave equation, and
calculate in detail the viscous coupling between the outer and inner cores and
the shell in two Appendices. These details appear not to have been published
previously. Explicit expressions are obtained for the reciprocal $Q$'s at both
boundaries, and for the viscous coupling torques between the outer core and the
shell (mantle plus crust), and between the outer and inner cores. The inner
core is found to be tightly coupled by viscosity to the outer core in the free
core nutations.

While the variational calculations of Jiang (1993) showed that there
should be a prograde free core nutation (PFCN), in addition to the classical
retrograde free core nutation (RFCN), he was anticipated in this discovery
by both Mathews et al (1991), and by de Vries and Wahr (1991), in studies of
the response of the Earth to nutational forcing. At the time, the VLBI nutation
record was too short to give an acceptable confidence level, and no claim of
observational support was made.

   There has been much interest in the detection of the inner core
translational triplet of modes in the spectra of superconducting gravimeter
observations since the initial identification fifteen years ago. They were
first identified visually in the Product Spectrum of a total of $110,000$ hours
of observations at four European stations (see Figure 9 of Smylie, Jiang,
Brennan and Sato, 1992). By adjustment of the degree one internal load Love
numbers, they were quickly shown to have the correct splitting to be the
translational triplet. A more complete study was made by Smylie, Hinderer,
Richter and Ducarme (1993), including a complete analysis of the statistics of
the Product Spectrum and a search across $4,119$ frequencies for correctly
split triplets in the subtidal band between $2 hr$ and $8hr$. The details
of this search are summarized in Section 3 of the present paper. It was shown
that the triplet originally identified visually had only one chance in
$6.8\times 10^{38}$ of being random. Nonetheless, the original identification
generated a flurry of papers, pro and con, on this result. Hinderer, Crossley
and Jensen (1995) were able to confirm the earlier result in the Product
Spectrum of the same data sets, although by a different spectral analysis
technique. Smylie, Hinderer, Richter and Ducarme (1993) had used the Welch
Overlapping Segment Analysis (WOSA) method, in which $12,000\;hr$ segments,
windowed with a Parzen window with $75\%$ overlap, enabled an improved signal
to noise ratio. In this case, the variance is inflated by a factor of $1.5$ but
the record length is quadrupled, for an overall increase in effective
record length of $8/3$. Hinderer, Crossley and Jensen (1995) were unable to
find the triplet in the cross spectrum of two years of common observations
at just two stations, Cantley, Canada and Strasbourg, France. They had hoped
for an improved result, even though simultaneous, short records from only two
stations were used, because the cross spectrum takes phase information into
account.

   Apart from the difficulties associated with the low signal to noise ratio of
the translational resonances in the spectra of superconducting gravimeter
observations, a second source of confusion is the calculation of the
translational mode periods for a given Earth model. Most calculations use
truncated vector spherical harmonic expansions derived from normal mode theory
(Rogister, 2003). The problem with this approach is that the Coriolis
acceleration couples these modes into slowly convergent infinite chains.
Typically, these infinite chains are truncated after only a few terms. Johnson
and Smylie (1977) showed that at periods of hours and more, convergence is very
slow and that many terms need to be included for accurate computation of
periods. This led Smylie, Jiang, Brennan and Sato (1992) to develop a
variational, finite element method of computing long period core modes using
bicubic splines as support functions. This method was used in the calculation
of translational mode periods and was extended by Jiang (1993) to determine
solutions for the free core nutations. The finite element method, using local
basis functions rather than global spherical vector harmonics, avoids the
Coriolis coupling problem and allows the accurate calculation of periods. The
initial formulation was criticized by Crossley, Rochester and Peng (1992) on
the grounds that dynamical terms arising from pure translations had been
neglected. This shortcoming was easily corrected (Smylie and Jiang, 1993). They
showed that the degree one reciprocal Love number for the inner core has a
quadratic dependence on frequency, as illustrated in Figure 5 of that paper.
This shortcoming did not affect the original identification of the
translational triplet, and current computations take the frequency dependence
of Love numbers, used to describe the deformations of the shell (mantle and
crust) and inner core, fully into account. The inviscid translational mode
periods for four Earth models are tabulated in Table 2, and are used in the
splitting law plot shown in Figure 6. The computation of these periods is now
done routinely to six or seven significant figures.

   Nonetheless, there remains a wide range of quoted translational triplet
periods in the literature. Crossley, Rochester and Peng (1992) give the triplet
for Earth model 1066A as $\left(4.04970,4.43806,4.89647\right)$ hours, close to
the known values in Table 2 of $\left(4.0491,4.4199,4.8603\right)$ hours. The
next year, Crossley (1993) found $\left(4.127,4.533,5.016\right)$ hours, while
Rogister (2003), using normal mode theory, got $\left(4.129,4.529,5.024\right)$
hours. Similarly, Rosat, Rogister, Crossley and Hinderer (2006) quote
$\left(4.64776,5.24395,5.83488\right)$ hours and $\left(4.74744,5.35865,5.92142
\right)$ for Earth model PREM, compared to the known triplet $\left(4.6776,
5.1814,5.7991\right)$ hours in Table 2. The noise level of quoted periods in
the literature, even among papers with common authors, seems to exceed the
noise level in the
spectra of superconducting gravimeter observations by a large margin! While the
scatter of reported periods may be due to attempts to apply truncated normal
mode theory, the situation was complicated, even further, by Rieutord (2002)
who finds $\left(3.894,4.255,4.687\right)$ hours for Earth model 1066A. He
misquotes the periods of Crossley, Rochester and Peng (1992) as $\left(4.95,
4.438,4.896\right)$ hours. The divergence of Rieutord's periods from known
values, apart from typographical errors, appears to be due to his assumption
that the shell (mantle and crust) and inner core are perfectly rigid, and that
the fluid outer core is perfectly adiabatically stratified. He further
dismisses the measurement of viscosity by Smylie and McMillan (2000), from the
reduction of rotational splitting of the two equatorial translational modes, on
the grounds that second order Ekman boundary layer theory is required, even
though he fails to perform the required second order analysis to support this
claim. For the observed Ekman number of $1.2\times 10^{-2}$, since the Ekman
layer theory is an ascending series in the square root of the Ekman number,
second order theory would give a correction of $0.11$ or $11\%$. The two
independent measures of viscosity from the two equatorial modes differ by
$9.6\%$. It is not known if this difference is due to the neglect of second
order terms in the Ekman boundary layer theory.  

   The most comprehensive analysis of superconducting gravimeter observations
appears to be that of Courtier et al (2000). A total of $294,106$ hours were
assembled for analysis. In addition to Product Spectra, a global multistation
experiment was performed using simultaneous observations from Brussels,
Cantley, Kakioka, Strasbourg and Wuhan, allowing the use of phase to lower
noise levels. Four independent sets of translational mode periods were
recovered, including those shown in Figure 3 of the present paper, all
differing from the original identification by no more than in the fifth
significant figure. A similar global experiment using the Product Spectrum was
attempted by Kroner, Jahr and Jentzsch (2004) for observations at five stations
without result. Insufficient detail is given to determine if the data sets
were correctly windowed to prevent frequency mixing from finite record effects,
but the presence of tidal lines in the subtidal band suggests not. Also the
inclusion of the Boulder data set, which has known serious timing errors, may
have compromised the results. Finally, little is known about the excitation of
the translational modes. Like the free core nutations, they may not be
continuously excited, and may not be present in more recent observations.

   By contrast, a recent new development in the search for the translational
modes has been achieved by Pagiatakis, Yin and El-Gelil (2007). Instead of
using the Product Spectrum based on $12,000$ hour segments, across many
stations, they use only the Cantley record of seventy-two months length. For
each month they compute a least squares periodogram. Then, in a variation of
the Product Spectrum method, they use the geometric mean of the seventy-two
periodograms to look for translational modes. While the use of only one month
data segments limits resolution, they find support for the periods found in
the original detection using the Product Spectrum.

   As illustrated in Figure 7, the translational mode periods are very
sensitive to inner core density. From Table 3, we see that the Cal8 Earth model
of Bolt and Uhrhammer gives periods very close to those observed. A reduction
of only $2.25\;milligrams\cdot cm^{-3}$ in inner core density brings the three
periods for this Earth model into coincidence with the observed triplet
(Smylie, Francis and Merriam, 2001). In terms of the density jump at the
inner core boundary, the Cal8 value of $1,170\;kg\cdot m^{-3}$ is reduced to
$1,168\;kg\cdot m^{-3}$.

   Traditionally, the dichotomy between extrapolated laboratory values of
viscosity and directly observed values for the Earth's core (Lumb and Aldridge,
1991), the latter being many orders of magnitude larger, has been explained by
dynamo theorists as due to turbulent flow, with the observed values reflecting
eddy viscosities. It is difficult to see how turbulent flow could explain
observed viscosities spanning nearly eight orders of magnitude across the core.
This view is also challenged by the close agreement of our
measured boundary values with the Arrhenius extrapolation of laboratory
experiments by Brazhkin (1998) and by Brazhkin and Lyapin (2000). The latter
reflect molecular viscosities. While turbulent flow may enhance dynamo action,
successful numerical dynamos with laminar flow, and with Ekman numbers in the
range implied by our measured viscosities have been routinely achieved
(Olsen and Christiansen, 2002).

\section{Viscosity at the Top of the Outer Core}

   In realistic Earth models, the finite element-based variational calculations
of Jiang (1993) showed that there are two free core nutations. In addition to
the classical retrograde mode (RFCN), a second prograde mode (PFCN) appears.
The motions are illustrated by Poinsot constructions, in which a small body
cone rolls once per sidereal day without slipping on a large space cone as
shown Figure 1 below.
\begin{center}
\begin{figure}[h]
\hspace*{3cm}
\epsfbox{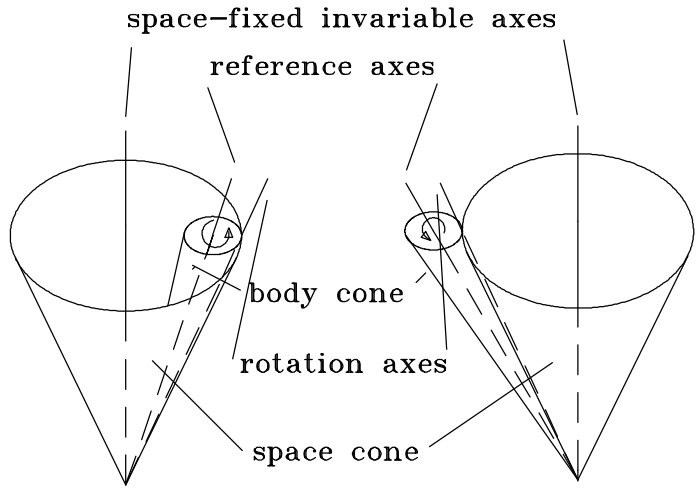}
\caption{Poinsot constructions for the Free Core Nutations. The RFCN is shown
on the left, the PFCN on the right. In each mode, the small body cone rolls
once per sidereal day without slipping on the large space cone. The line of
contact is Earth's instantaneous rotation axis.}
\end{figure}
\end{center}
 
   In the Earth frame, both modes appear as nearly diurnal retrograde wobbles,
the wobble associated with the RFCN is a little faster than retrograde diurnal,
while the wobble associated with the PFCN is a little slower. For the RFCN, the
ratio of the nutation amplitude $a_N$ to the wobble amplitude $a_W$ is given by
$a_N/a_W-1=-\Omega/\sigma_N$, where $\Omega$ is Earth's angular rotation
velocity and $\sigma_N$ is the nutation angular frequency. For the PFCN, the
ratio is given by $a_N/a_W +1=\Omega/\sigma_N$.

   In the study of Palmer and Smylie (Palmer and Smylie, 2005), both the
RFCN and the PFCN were found to be in free decay, on the basis of VLBI nutation
measurement series from GSFC and the USNO in excess of twenty-three years
length. The GSFC series runs from August 3, 1979 to March 6, 2003 and is taken
to span a period of 8,617 days, while the USNO series runs from August 3, 1979
to March 29, 2003 and is taken to span 8,631 days. In order to investigate the
decays in detail, each record was divided into 2,000 day segments, advancing
down the time axis in 400 day steps. Spectral densities were estimated on the
basis of each of four successive 2,000 day segments. The result was spectral
estimates centered at 1,600 days into the record, and at 400 day increments
down the time axis thereafter. The spectral amplitudes are shown plotted
against time in Figure 2.
\begin{center}
\begin{figure}[h]
\hspace*{2cm}
\epsfbox{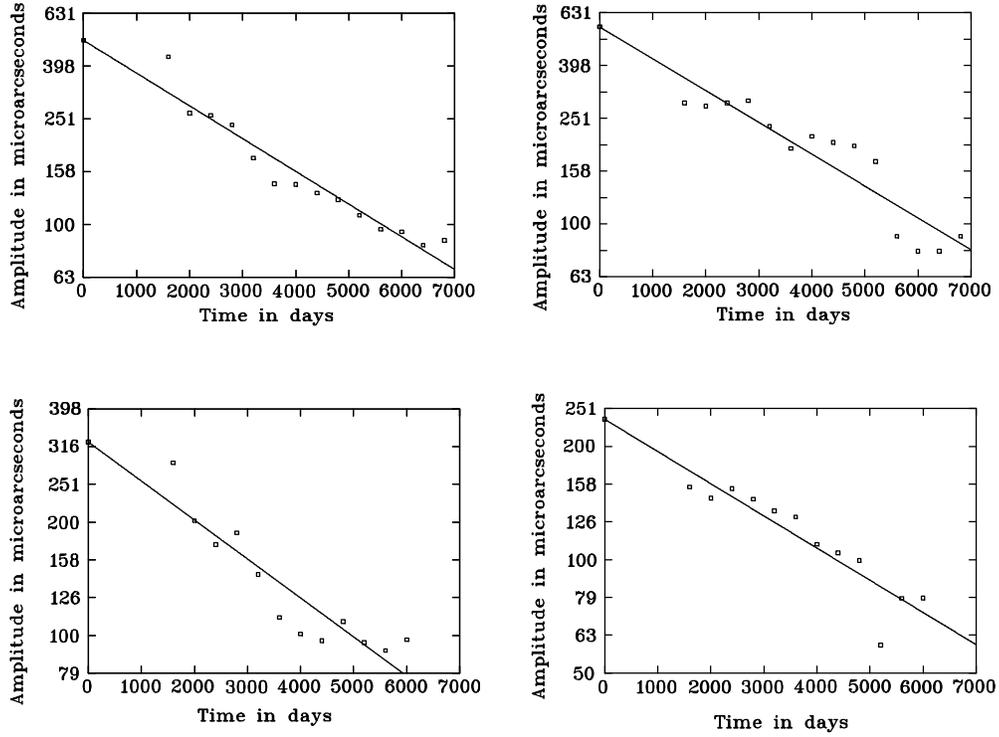}
\caption{Amplitudes of the free core nutations plotted on a logarithmic scale
as functions of time. Upper plots show the RFCN, upper left for GSFC, upper
right for USNO, lower plots show the PFCN, lower left for GSFC, lower right for
USNO. Linear fits to the time dependencies are shown directly on the plots,
along with extrapolated values at the time origin.}
\end{figure}
\end{center}

   In free decay, the logarithm of the nutation amplitude $a_N$ decreases
linearly with time and
\bequ\label{eqn34}
\log a_N=ct+d,
\nequ
with
\begin{align}
&c=\pm\frac{\pi\log e}{Q_N T_N}=-\frac{\log e}{\tau},\;\;d=\log a_{N_0},
\nonumber\\
&t_{1/2}=\tau\ln 2,\\
\end{align}
where $T_N$ is the signed nutation period (negative for retrograde, positive
for prograde), $Q_N$ is the apparent $Q$ of the nutation in the space frame,
$a_{N_0}$ is the amplitude at time $t=0$, $\tau$ is the e-folding time and
$t_{1/2}$ is the half life of the decay. The upper positive sign applies to
the RFCN, while the lower negative sign applies to the PFCN.

   The actual physical dissipation takes place in the Earth frame, through the
associated nearly diurnal retrograde wobbles, and is measured by $Q_W$. $Q_W$
is related to the apparent nutation $Q$ by
\bequ
Q_W=\pm\left(1-T_N/T_s\right)Q_N,
\nequ
with $T_s$ being the length of the sidereal day. The wobble $Q$ is found to be
\bequ
Q_W=-\frac{\pi\log e}{cT_s}\left(1-T_s/T_N\right).
\nequ  
Results of the regression on the logarithm of the nutation amplitude,
(\ref{eqn34}), are given in Table 1.
\begin{table}[h]
\centering
\caption{Fitted decay parameters of the Free Core Nutations.}
\begin{tabular}{llccccccc}
\hline
&&$c\left(10^{-4}days^{-1}\right)$&$d$&$a_{N_0}\left(\mu as\right)$&$T_N
\left(days\right)$&$Q_N$&$Q_W$&$t_{1/2}
\left(days \right)$\\
\hline
RFCN&&&&&&\\
&GSFC&$-1.23909$&$2.69740$&$498.20$&$-440.865$&$26.404$&$11,068$&$2,429$\\
&&$\pm 0.1006$&&&$\pm 31.519$&$\pm 4.031$&$\pm 2,481$&\\
&USNO&$-1.20360$&$2.74547$&$556.51$&$-410.147$&$27.444$&$11,394$&$2,501$\\
&&$\pm 0.1486$&&&$\pm 17.706$&$\pm 4.573$&$\pm 2,390$&\\
PFCN&&&&&&\\
&GSFC&$-1.03217$&$2.51192$&$325.03$&$474.308$&$34.052$&$13,221$&$2,916$\\
&&$\pm 0.1461$&&&$\pm 86.578$&$\pm 11.036$&$\pm 6,698$&\\
&USNO&$-0.85237$&$2.37147$&$235.22$&$444.847$&$41.356$&$16,009$&$3,532$\\
&&$\pm 0.1236$&&&$\pm 74.677$&$\pm 12.939$&$\pm 7,696$\\
\hline
\end{tabular}
\end{table}

   In realistic Earth models, the two free core nutations have associated
nearly diurnal retrograde wobbles that are close to pure rotations with
respect to the shell (mantle and crust) and the inner core (Jiang, 1993;
Jiang and Smylie, 1995; Jiang and Smylie, 1996). Their angular frequencies
are close to $-\Omega$, the negative of Earth's angular rotation velocity.
In spherical polar co-ordinates $\left(r,\theta,\phi\right)$ the velocity
field is closely
\bequ\label{eqn4}
{\bm v}=-\hat{\bm \theta}Ar\sin\left(\phi +\Omega t\right)-\hat{\bm \phi}
Ar\cos\theta\cos\left(\phi +\Omega t\right),
\nequ
for nearly diurnal wobble amplitude A.

   On the assumption that the boundary layers are a small fraction of the
radius in thickness, the leading order boundary layer equations in the
$\left(\theta,\phi\right)$ components of the extra boundary layer velocity,
$\left(v_\theta,v_\phi\right)$, are (Moore, 1978; Smylie and McMillan, 1998)
\begin{align}
\frac{\partial v_\theta}{\partial t}-2\Omega v_\phi\cos\theta=&\nu
\frac{\partial^2 v_\theta}{\partial r^2},\nonumber\\
\frac{\partial v_\phi}{\partial t}+2\Omega v_\theta\cos\theta=&\nu\frac{
\partial^2 v_\phi}{\partial r^2},\label{eqn6}
\end{align}
with $\nu$ denoting the kinematic viscosity.

   The detailed solution of these boundary layer equations, and the calculation
of the rates of energy dissipation in each of the respective layers at the
boundaries of the outer core is left to Appendix A. From equation
(\ref{eqn24}) of that Appendix, the rate of energy
dissipation in each of the respective boundary layers is found to be
\bequ
\frac{dE}{dt}=\frac{\pi}{35}\rho_0 A^2 r_0^4\sqrt{2\nu\Omega}\left(9\sqrt{3}
+19\right)
\nequ
with $\rho_0$ denoting the density, $A$ the amplitude, and $r_0$ the radius
at the boundaries. If $A_a$ is the amplitude of the nearly diurnal retrograde
wobble of the outer core with respect to the inner core, and if $A_b$ is the
amplitude with respect to the shell, the total energy dissipated per cycle
in both boundary layers is
\bequ
E=\frac{2\pi}{\Omega}\cdot\frac{dE}{dt}=\frac{2}{35}\pi^2\left(\rho_0\left(a
\right)A_a^2 a^4\sqrt{\nu_a}+\rho_0\left(b\right)A_b^2 b^4\sqrt{\nu_b}\right)
\sqrt{\frac{2}{\Omega}}\left(9\sqrt{3}+19\right),
\nequ
with $\rho_0\left(a\right)$ and $\nu_a$ representing the density and kinematic
viscosity just outside the ICB (at $r_0=a$) and $\rho_0\left(b\right)$ and
$\nu_b$ representing the density and kinematic viscosity just inside the CMB
(at $r_0=b$). The quality factor of the nearly diurnal wobbles accompanying the
free core nutations, $Q_W$, is defined as $2\pi$ times the ratio of the total
energy to the energy dissipated per cycle, $E$. The total energy of the motion
is closely
\bequ
\frac{1}{2}I_{oc}A^2,
\nequ
where $I_{oc}=911.79\times 10^{34}\;kg\cdot m^2$ is the moment of inertia of
the outer core. The reciprocal of the overall quality factor is then the sum
of the reciprocals of the effective quality factors at the two boundaries,
\bequ\label{eqn20}
\frac{1}{Q_W}=\frac{1}{Q_a}+\frac{1}{Q_b},
\nequ
with
\bequ
\frac{1}{Q_a}=\frac{2\pi\rho_0\left(a\right)a^4\sqrt{2\nu_a/\Omega}\left(9
\sqrt{3}+19\right)}{35I_{oc}},
\nequ
\bequ
\frac{1}{Q_b}=\frac{2\pi\rho_0\left(b\right)b^4\sqrt{2\nu_b/\Omega}\left(9
\sqrt{3}+19\right)}{35I_{oc}}.
\nequ

Neglecting the perturbation on shell rotation, if the amplitude of the nearly
diurnal wobble of the outer core is $B$ and that of the inner core is $C$,
\[A_b=B,\;\;\;A_a=B-C.\]

The detailed calculation of the viscous coupling torques the outer core exerts
on the shell and inner core is left to Appendix B. From expressions
(\ref{eqn30}) and (\ref{eqn31}) of that Appendix, and ignoring the small
Chandler resonance effect, the equation of motion of the outer core is
\bequ\label{eqn28}
\gamma_a e^{-i\Omega t}\left(C-B\right)-\gamma_b e^{-i\Omega t}B=I_{oc}\left(
\dot{B}-i\Omega B\right)e^{-i\Omega t},
\nequ
while that of the inner core is
\bequ\label{eqn29}
-\gamma_a e^{-i\Omega t}\left(C-B\right)=I_{ic}\left(\dot{C}-i\Omega C\right)
e^{-i\Omega t},
\nequ
with $I_{ic}$ representing the moment of inertia of the inner core.
Equations (\ref{eqn28}) and (\ref{eqn29}) constitute a linear, homogeneous
differential system. For time dependence $e^{\lambda t}$, it becomes
\bequ\label{eqn32}
\left(\begin{array}{cc}
\lambda -i\Omega +\left(\gamma_a +\gamma_b\right)/I_{oc}&-\gamma_a/I_{oc}\\
-\gamma_a/I_{ic}&\lambda -i\Omega +\gamma_a/I_{ic}
\end{array}\right)\left(\begin{array}{c}B\\C\end{array}\right)=0,
\nequ
with characteristic equation
\bequ
\lambda^{\prime 2} +\left(\frac{1+\gamma_b/\gamma_a}{I_{oc}}+\frac{1}{I_{ic}}
\right)\gamma_a \lambda^\prime +\frac{\gamma_a\gamma_b}{I_{oc}I_{ic}}=0,
\nequ
where $\lambda^\prime=\lambda -i\Omega$.

   From (\ref{eqn30}), (\ref{eqn31}), the ratio
\bequ
\frac{\gamma_b}{\gamma_a}=\frac{\rho_0\left(b\right)b^4\sqrt{\nu_b}}{\rho_0
\left(a\right)a^4\sqrt{\nu_a}}=1/117.4,
\nequ
for density ratio $0.8069$, radius ratio $2.8668$, and the ratio of the square
roots of the kinematic viscosities $1/6400$. Correct to terms of first order
in the small quantity $\gamma_b/\gamma_a$, the roots of the characteristic
equation are
\bequ
\lambda^\prime_1=\frac{\gamma_b}{I_c},\;\mbox{and}\;\lambda^\prime_2=-
\frac{I_c}{I_{oc}I_{ic}}\gamma_a -\frac{I_{ic}}{I_{oc}I_c}\gamma_b,
\nequ
with $I_c=I_{oc}+I_{ic}$ representing the moment of inertia of the entire core.
The admissible solutions of the system (\ref{eqn32}) are then the linear
combinations
\begin{align}
Be^{-i\Omega t}=&\alpha e^{\lambda_1^\prime t}+\beta e^{\lambda_2^\prime t},
\nonumber\\
Ce^{-i\Omega t}=&\left(1+\frac{\gamma_b}{\gamma_a}\frac{I_{ic}}{I_c}\right)
\alpha e^{\lambda_1^\prime t}-\left(\frac{I_{oc}}{I_{ic}}-\frac{\gamma_b}
{\gamma_a}\frac{I_{oc}}{I_c}\right)\beta e^{\lambda_2^\prime t},
\end{align}
with $\alpha$, $\beta$ representing arbitrary linear combination coefficients.
The decay times of the two solutions are, respectively
\begin{align}
\tau_1=&\frac{I_c}{Rl\gamma_b},\\
\tau_2=&\frac{I_{oc}I_{ic}}{\left(I_c Rl\gamma_a +I_{ic}Rl\gamma_b\right)},
\end{align}
where
\begin{align}
Rl\gamma_a=&\pi\rho_0\left(a\right)a^4\sqrt{\nu_a\Omega}\frac{\sqrt{2}}{35}
\left[9\sqrt{3}+19\right],\\
Rl\gamma_b=&\pi\rho_0\left(b\right)b^4\sqrt{\nu_b\Omega}\frac{\sqrt{2}}{35}
\left[9\sqrt{3}+19\right].
\end{align}
Again, correct to first order in the small quantity $\gamma_b/\gamma_a$, the
ratio of the decay times is
\bequ
\frac{\tau_2}{\tau_1}=\frac{I_{oc}I_{ic}}{I_c^2}\frac{Rl\gamma_b}{Rl\gamma_a}=
\frac{I_{oc}I_{ic}}{I_c^2}\frac{\gamma_b}{\gamma_a}.
\nequ
With $I_{oc}=911.79\times 10^{34}\;kg\cdot m^2$, $I_{ic}=6.16\times 10^{34}\;
kg\cdot m^2$ and, hence $I_c=917.95\times 10^{34}\;kg\cdot m^2$, we have
$\tau_2/\tau_1=1/17,613$. Thus, the second solution damps out rapidly, and we
are left with
\bequ
C=\left(1+\frac{\gamma_b}{\gamma_a}\frac{I_{ic}}{I_c}\right)B,
\nequ
and
\bequ
B-C=-\frac{\gamma_b}{\gamma_a}\frac{I_{ic}}{I_c}B.
\nequ

   From expression (\ref{eqn24}), the total rate of energy dissipation at both
boundaries then becomes
\bequ
\frac{dE}{dt}=\pi\rho_0\left(b\right)A_b^2 b^4\sqrt{\nu_b\Omega}\frac{\sqrt{2}}
{35}\left(9\sqrt{3}+19\right)\left[1+\frac{\gamma_b}{\gamma_a}\left(\frac{
I_{ic}}{I_c}\right)^2\right].
\nequ
The energy dissipated per cycle is
\bequ
E=\frac{2\pi}{\Omega}\cdot\frac{dE}{dt}=\frac{2}{35}\pi^2\rho_0\left(b\right)
A_b^2 b^4\sqrt{\frac{2\nu_b}{\Omega}}\left(9\sqrt{3}+19\right)\left[1+
\frac{\gamma_b}{\gamma_a}\left(\frac{I_{ic}}{I_c}\right)^2\right].
\nequ
The total energy of the motion is
\bequ
T=\frac{1}{2}I_{oc}A_b^2 +\frac{1}{2}I_{ic}\left[1+\frac{\gamma_b}{\gamma_a}
\frac{I_{ic}}{I_c}\right]^2=\frac{1}{2}I_c\left[1+2\frac{\gamma_b}{\gamma_a}
\left(\frac{I_{ic}}{I_c}\right)^2\right]A_b^2
\nequ
to first order in the small quantity $\gamma_b/\gamma_a$. To the same order,
the wobble quality factor is
\bequ
Q_W=2\pi T/E=\frac{35I_c\left[1+\gamma_b/\gamma_a\left(I_{ic}/I_c\right)^2
\right]}{2\pi\rho_0\left(b\right)b^4\sqrt{2\nu_b/\Omega}\left(9\sqrt{3}+19
\right)}.
\nequ
Finally, correct to first order, the viscosity recovered from the
observed $Q_W$ of the nearly diurnal retrograde wobbles is
\bequ\label{eqn33}
\nu_b=\frac{1225I_c^2\Omega\left[1+2\gamma_b/\gamma_a\left(I_{ic}/I_c\right)^2
\right]}{8\pi^2\rho_0^2\left(b\right)b^8\left(9\sqrt{3}+19\right)^2 Q_W^2}.
\nequ
We see from this expression that the inner core is tightly coupled to the
outer core motion, and that the correction for dissipation in the lower
boundary layer, represented by the quantity in square brackets on the
numerator, differs from unity by only $3.8\times 10^{-7}$ and may be neglected.

   Using expression (\ref{eqn33}), and the values of $Q_W$ listed in Table 1,
we recover a viscosity of $3,038\pm1,362\;Pa\cdot s$ for the RFCN from the
GSFC series, $2,866\pm1,203\;Pa\cdot s$ from the USNO series,
$2,129\pm 2,157\;Pa\cdot s$ for the PFCN from the GSFC series and $1,452
\pm 1,396\;Pa\cdot s$ from the USNO series. The mean value of the recovered
dynamic viscosity is then $2,371\pm 1,530\;Pa\cdot s$. For a density of
$9.82\times 10^3\;kg\cdot m^{-3}$ at the top of the core, the corresponding
kinematic viscosity is $0.2414\pm 0.1558\;m^2 s^{-1}$.

\section{Viscosity at the Bottom of the Outer Core}

  Near the bottom of the outer core, the viscosity in the F-layer
(Smylie, 1999; Smylie and McMillan, 2000) can be found from the reduction in
the rotational splitting of the two equatorial translational modes of the Inner
Core. The translational modes are observed in the Product Spectrum of global
networks of superconducting gravimeters (Smylie, Hinderer, Richter and Ducarme,
1993; Courtier et al., 2000). In Figure 3, we show the three translational mode
resonances found in the Product Spectrum, based on observations at Bad Homburg
(24,272 hours), Brussels (83,892 hours), Cantley (32,992 hours) and Strasbourg
(78,504 hours). Both the prograde and axial mode resonances are well above the
95\% C.I. and the retrograde mode is just below this level of significance
(Fig.8 (b), Courtier et al., 2000).
\begin{center}
\begin{figure}[h]
\epsfbox{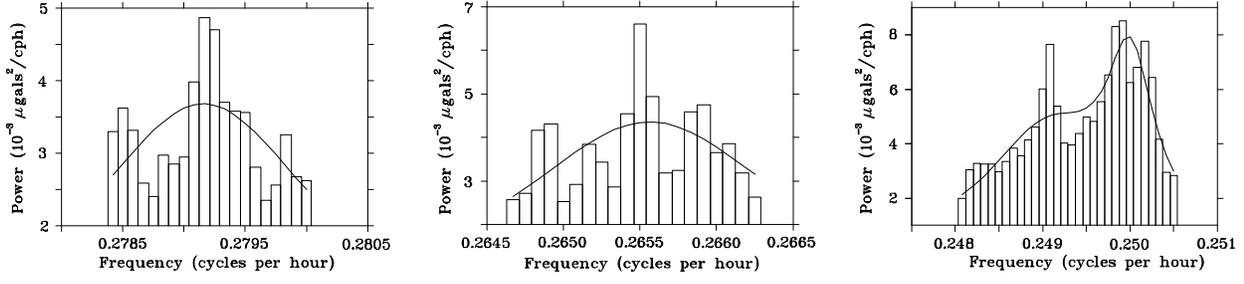}
\caption{Product Spectra of (from left to right) the retrograde equatorial,
axial and prograde equatorial translational modes of the Inner Core. The
prograde equatorial mode is near the large solar heating tide feature $S_6$
at exactly six cycles per solar day. The recovered central periods are,
respectively $3.5822\pm0.0012\;hr$, $3.7656\pm0.0015\;hr$ and $4.0150\pm
0.0010\;hr$.}
\end{figure}
\end{center}

   A much more stringent test of significance arises from consideration of the
pressure and viscous drags on the Inner Core (Smylie and McMillan, 2000) which
leads to a splitting law of the form
\bequ\label{eqn50}
\left(\frac{T}{T_0}\right)^2 +2g^\nu\frac{T_0}{T_s}\left(\frac{T}{T_0}\right)-
1=0,
\nequ
where $T$ is the period, $T_0$ is the unsplit period, $T_s$ is the length of
the sidereal day, and $g^\nu$ is a dimensionless viscous splitting parameter.
For the axial mode, the viscous splitting parameter is related to the inviscid
splitting parameter $g^i$ by
\bequ
g^\nu =g^i\left[1+\frac{1}{4}\frac{M_I -M_I^\prime}{M_I +\alpha}\sqrt{E_k}
f^a\left(\sigma\right)\right],
\nequ
and for the equatorial modes by
\bequ
g^\nu =g^i\left[1-\frac{1}{8}\left(\frac{M_I^\prime -\beta}{M_I +\beta}+
\frac{M_I^\prime +\alpha}{M_I +\alpha}\right)\sqrt{E_k}f^e\left(\sigma\right)
\right].
\nequ
$\alpha$, $\beta$ are coefficients of the pressure drag on the Inner Core
given by
\bequ
\alpha=M_I^\prime\left(\frac{1}{2}+\frac{3}{2}\frac{M_I +\left(a/b\right)^3
M_S}{M_O +M_S\left(1-\left(a/b\right)^3\right)}\right),
\nequ
and
\bequ
\beta =M_I^\prime\left(\frac{1}{4}-\frac{3}{4}\frac{M_I +\left(a/b\right)^3
M_S}{M_O +M_S\left(1-\left(a/b\right)^3\right)}\right).
\nequ
$M_I$ is the mass of the Inner Core, $M_O$ is the mass of the outer core,
$M_S$ is the mass of the shell, and $M_I^\prime =4/3\pi a^3\rho_0\left(a
\right)$ is the displaced mass, $\rho_0\left(a\right)$ the density at
the bottom of the outer core. $\sigma=\omega/2\Omega$ is the dimensionless
Coriolis frequency corresponding to angular frequency $\omega$. $f^a\left(
\sigma\right)$, $f^e\left(\sigma\right)$ are dimensionless functions of
$\sigma$ given by
\bequ
f^a\left(\sigma\right)=\left\{8\left[\left(\sigma +1\right)^{3/2}+\left(\sigma 
-1\right)^{3/2}\right]-\frac{16}{5}\left[\left(\sigma +1\right)^{5/2}-\left(
\sigma -1\right)^{5/2}\right]\right\},
\nequ
and
\bequ
f^e \left(\sigma\right)=\left\{\mp 24\left(\pm\sigma\mp1\right)^{1/2} -16\left(
\pm\sigma\mp 1\right)^{3/2} -\frac{16}{5}\left[\left(\pm\sigma -1\right)^{5/2}
-\left(\pm\sigma +1\right)^{5/2}\right]\right\},
\nequ
with the upper sign referring to the retrograde mode, for which $\sigma$ is
positive and the lower sign referring to the prograde mode, for which $\sigma$
is negative.

   For three candidate periods, $T_R$ (retrograde), $T_C$ (axial) and $T_P$
(prograde), the splitting equation (\ref{eqn50}) provides the corresponding
values of the dimensionless viscous splitting parameter, $g_R^\nu$, $g_C^\nu$,
and $g_P^\nu$, for a given value of $T_0$. Thus, the whole frequency axis can
be searched for correctly split resonances. For a resonance centred on
frequency $f_j$, its form at neighbouring frequencies $f_i$ is
\bequ\label{eqn51}
r_{ij}=\frac{a_j^2}{1+4Q\left[\left(f_i -f_j\right)/f_j\right]^2}.
\nequ
For record segments of $12,000\;hour$ length, Product Spectral estimates
$s_i$ are spaced at intervals of $1/12,000\;cycles/hour$ along the
frequency axis. In the subtidal band, between $2\;hr$ and $8\;hr$ period,
there are $4,501$ spectral estimates. For twenty-five spectral estimates
centred on frequency $f_j$ with $Q=100$, the misfit of (\ref{eqn51}) to
spectral estimate $s_i$ is
\bequ
\epsilon_{ij}=A_j r_{ij}-s_i.
\nequ
The error energy of the fit is
\bequ
I_j=\sum_{i=j-12}^{j+12}\epsilon_{ij}^2.
\nequ
Minimizing the error energy of the fit gives
\bequ
A_j =\sum_{i=j-12}^{j+12}r_{ij}s_i /\sum_{i=j-12}^{j+12}r_{ij}^2,
\nequ
with minimum error energy
\bequ
I_{min}=\sum_{i=j-12}^{j+12}s_i^2 -A_j^2\sum_{i=j-12}^{j+12}r_{ij}^2.
\nequ
As a measure of the strength of a potential resonance of the form
(\ref{eqn51}),we use the parameter $S_j^2=A_j^2/I_{min}$. When a large well-fit
resonance is found, we expect $S_j^2$ to be large, and if a small, poorly fit
spectral feature is found, we expect $S_j^2$ to be small. For each of the
available $4,477$ frequencies, $f_j$, in the subtidal band, we set $T_0=1/f_j$
and compute $f_R=1/T_R$, $f_C=1/T_C$ and $f_P=-1/T_P$ from equation
(\ref{eqn50}). The values of $S_R^2$, $S_C^2$ and $S_P^2$ of the resonance
parameter $S^2$ at the discrete frequencies nearest $f_R$, $f_C$ and $f_P$,
respectively, are then multiplied together to form the splitting product $P_j$
as an indicator of the presence of correctly split resonances. We show the
resulting probability density function (PDF) for the splitting product computed
at $4,119$ points along the frequency axis in the subtidal band in
Figure 4.  
\begin{center}
\begin{figure}[h]
\hspace*{3cm}
\epsfbox{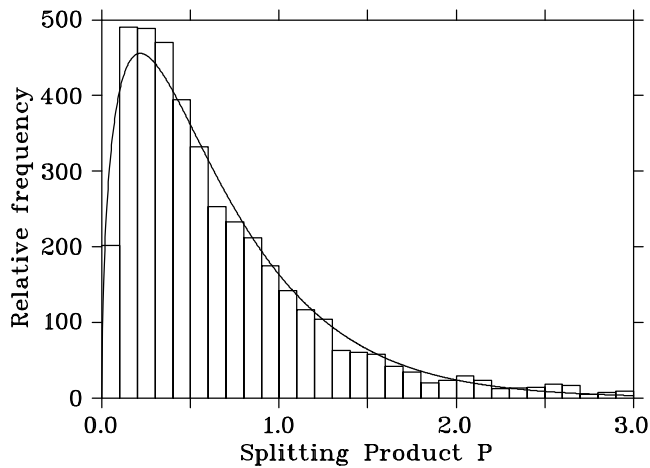}
\caption{Probability density function (PDF) for the splitting product $P$.
Bins in $P$ are $0.1$ wide. The fitted PDF is for a $\chi_\nu^2$ distribution
with $\nu=2.97614$ for the random variable $4.5314P$.}
\end{figure}
\end{center}

   The PDF shown in Figure 4 allows the evaluation of the significance of
translational triplets along the frequency axis. In Figure 5 we show the
splitting products found between $2\;hr$ and $10\;hr$ periods. A very large
value of $P$ is found at $T_0=3.7975\;hr$. From the PDF it is found that the
probability of a realization of $P_j$ larger than the largest shown in Figure
5 is only $1$ in $6.8\times 10^{38}$! The resonances shown in Figure 3 seem
to have been correctly identified as translational modes.
\begin{center}
\begin{figure}[h]
\hspace*{3cm}
\epsfbox{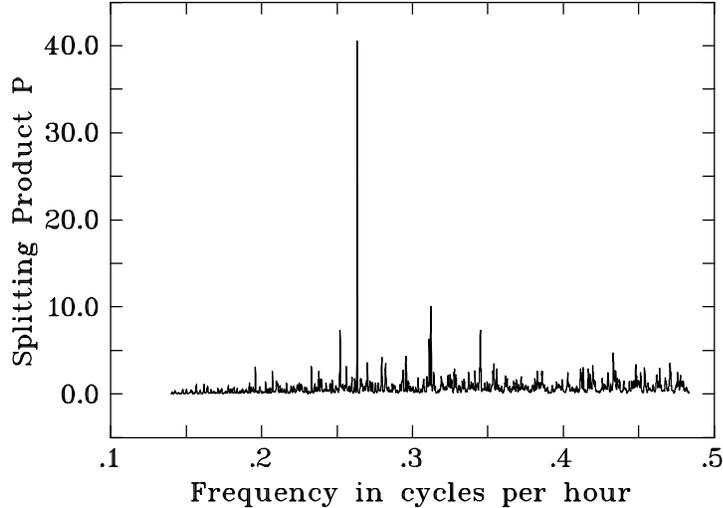}
\caption{Splitting product $P$ as a function of frequency. The large spike at
$T_0=3.7975$ corresponds to the translational triplet plotted in Figure3.}
\end{figure}
\end{center} 

   Plots of the splitting law for the three translational modes are shown in
Figure 6. {\em The inviscid periods for the four Earth models plotted in this
Figure are listed in Table 2.}
{\em
\begin{table}[h]
\centering
\caption{Inviscid periods for the four Earth models shown plotted in Figure 6.}
\begin{tabular}{llll}
\hline
Periods&Retrograde&Axial&Prograde\\
&$\left(hours\right)$&$\left(hours\right)$&$\left(hours\right)$\\
\hline
Core11 Periods&5.1280&5.7412&6.5114\\
PREM Periods&4.6776&5.1814&5.7991\\
1066A Periods&4.0491&4.4199&4.8603\\
Cal8 Periods&3.5168&3.7926&4.1118\\
\hline
\end{tabular}
\end{table} 
}

\begin{center}
\begin{figure}[h]
\hspace*{2cm}
\epsfbox{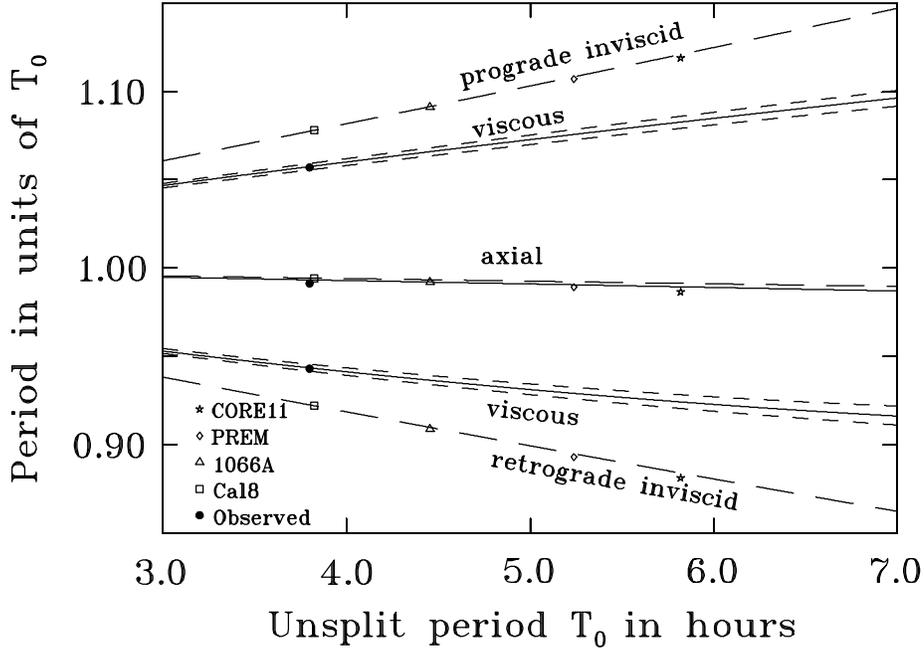}
\caption{Splitting curves for the three translational modes. The inviscid
curves are shown dashed using the splitting parameters for Earth model Cal8
(open squares) of Bolt and Uhrhammer (Bullen and Bolt, 1985, Appendix).
Inviscid periods are over plotted for Earth models Core11 (open stars)
(Widmer et al., 1988), PREM (open diamonds) (Dziewonski and Anderson, 1981) and
1066A (open triangles) (Gilbert and Dziewonski, 1975). Solid viscous splitting
curves are for a single viscosity of $1.247\times 10^{11}\;Pa\cdot s$.}
\end{figure}
\end{center}

   Two independent measures of viscosity are given, as the reduction in
rotational splitting is larger for the retrograde mode than for the prograde
equatorial mode. The retrograde equatorial mode gives
$1.190\pm0.035\times 10^{11}\;Pa\cdot s$, while the prograde equatorial mode
gives $1.304\pm0.034\times 10^{11}\;Pa\cdot s$. A balanced error value of
$1.247\times 10^{11}\;Pa\cdot s$ yields viscous periods that are only $6.5\;s$
longer than the observed periods.

   From Figure 6, we see that the observed periods are close to those for the
Cal8 Earth model. In Table 3 a detailed comparison of the Cal8 periods and
those observed is shown.
\begin{table}[h]
\centering
\caption{Comparison of the observed translational mode periods with those of
the Cal8 Earth model.}
\begin{tabular}{llll}
\hline
Periods&Retrograde&Axial&Prograde\\
&$\left(hours\right)$&$\left(hours\right)$&$\left(hours\right)$\\
\hline
Observed Periods&3.5822&3.7656&4.0150\\
Cal8 Viscous Periods&3.5840&3.7731&4.0168\\
Cal8 Inviscid Periods&3.5168&3.7926&4.1118\\
\hline
\end{tabular}
\end{table}

   The close match of the observed periods to those of the Cal8 Earth model is
due to the sensitivity of the translational mode periods to inner core density.
In Figure 7, we show the density profiles of the Inner Core for Earth models
Cal8, 1066A, PREM and Core11 together with their unsplit periods, $T_0$.
\begin{center}
\begin{figure}[h]
\hspace*{3cm}
\epsfbox{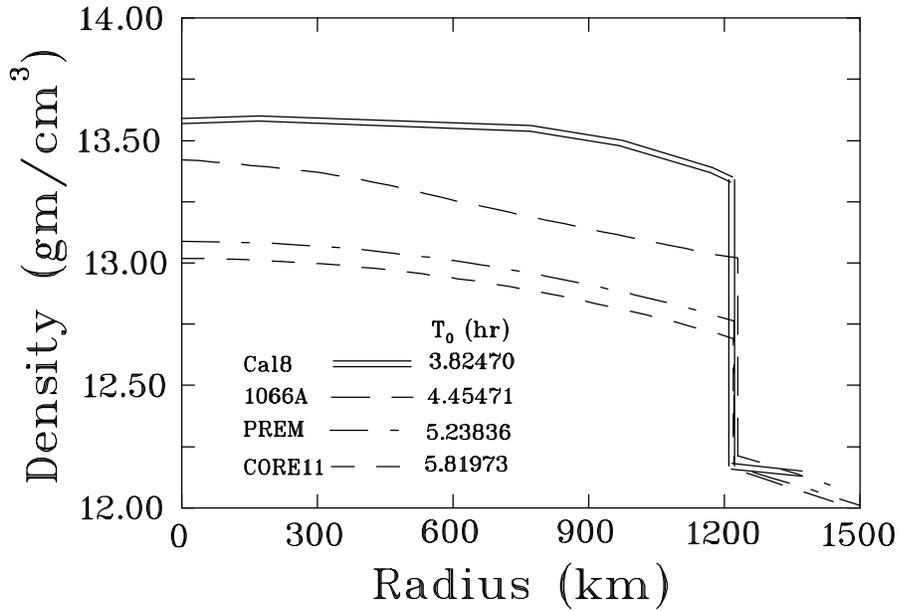}
\caption{Detailed density profiles of the Inner Core for Earth models Cal8,
1066A, PREM and Core11. The range of $0.6\;gm\cdot cm^{-3}$ causes a nearly
$2\;hr$ difference in the unsplit period, $T_0$, giving a resolution of
$200\;minutes/gm\cdot cm^{-3}$.}
\end{figure}
\end{center}
The axial mode period suffers little rotational or viscous splitting. Its
observed period provides a strict constraint on inner core density. The
calculated axial mode period for Cal8 is only $27\;s$ longer than the observed
period. An overall density decrease in the inner core of only $2.25\;milligrams
\cdot cm^{-3}$ would bring them into coincidence, giving a very strong
confirmation of Cal8. In Table 4 we show a comparison of the unsplit period,
$T_0$, for the four Earth models with the observed value.
\begin{table}[h]
\centering
\caption{Comparison of the unsplit period, $T_0$, for four Earth models with
the observed value.}
\begin{tabular}{llll}
\hline
Earth Model&Unsplit Period&Deviation&Error\\
&$T_0\left(hrs.\right)$&$\Delta T_0\left(hrs.\right)$&\%\\
\hline
Observed&$3.7985$&&\\
Cal8&$3.82470$&$0.0262$&$0.69$\\
1066A&$4.45471$&$0.65621$&$17.28$\\
PREM&$5.23836$&$1.43986$&$37.91$\\
Core11&$5.81973$&$2.02123$&$53.21$\\
\hline
\end{tabular}
\end{table}  

\section{A Viscosity Profile for the Outer Core}

   The boundary values of viscosity we have found are in very close agreement
with an Arrhenius extrapolation of their laboratory experiments by Brazhkin
(Brazhkin, 1998) and by Brazhkin and Lyapin (Brazhkin and Lyapin, 2000), who
find $10^{11}\;Pa\cdot s$ at the bottom of the outer core and $10^2\;Pa\cdot s$
at the top. We are prompted by the very close agreement of the viscosity
measures at the boundaries with those provided by the Arrhenius extrapolation,
to extend it to interpolate between the boundary values, to obtain a viscosity
profile across the entire outer liquid core. 
 
   The Arrhenius description of the temperature and pressure dependence of
the dynamic viscosity $\eta$ is (Brazhkin, 1998)
\bequ\label{eqn35}
\eta\sim\exp\left(\frac{E_{act_0}+PV_{act}}{kT}\right),
\nequ
with $E_{act_0}$ representing the activation energy at normal pressure, $P$ the
pressure, $V_{act}$ the activation volume, $k$ Boltzmann's constant and $T$ the
Kelvin temperature. $V_{act}$ is proportional to the atomic volume, which,
in turn, is inversely proportional to the density $\rho$. While the activation
volume for liquid metals at atmospheric pressure is very small, Brazhkin (1998)
and Brazhkin and Lyapin (2000) report experimental results on pure iron at the
melting temperature, $T_m$, that show it to be strongly pressure dependent up
to pressures of 95 kbar. The strong pressure dependence requires integration of
the differential form of the Arrhenius expression. For dominant pressure
dependence, from expression (\ref{eqn35}) the differential increment in
viscosity is proportional to
\bequ\label{eqn36}
\frac{D}{\rho T_m}\exp{D\frac{P}{\rho T_m}}dP,  
\nequ
with $D$ a pressure-dependent parameter, allowing for the pressure dependence
of the activation volume, and $dP$ being the differential increment in
pressure. The integral of (\ref{eqn36}) over pressure is easily converted to an
integral over radius $r$ since $dP/dr=-\rho g$, where $g$ is the gravitational
acceleration at radius $r$. The viscosity at radius $r$ is then
\bequ\label{eqn37}
\eta\left(r\right)=\eta_b+\eta_b\int_b^r\frac{D}{\rho T_m}\exp{\left(D\frac{P}
{\rho T_m}\right)}\frac{dP}{dr}dr,
\nequ
with $b$ the radius of the core-mantle boundary and $\eta_b=2,371\;Pa\cdot s$
the dynamic viscosity at the top of the core. To perform the integration in
(\ref{eqn37}), we require profiles of pressure, density, melting temperature
and pressure gradient. The pressure profile can be found by integrating the
product of gravity and density for an Earth model (we use Cal8, see Bullen and
Bolt (1985), p.472). The melting
temperatures are found by spline interpolation onto the Cal8 radii from those
tabulated by Stacey (1992, p.459). The required profiles are shown in Table 5.
\begin{table}[h]      
\centering
\caption{Pressure, density, melting temperature and radial pressure gradient
profiles.}
\begin{tabular}{lllll}
\noalign{\hrule height 1.5pt}
radius&$P$&$\rho$&$T_m$&$dP/dr$\\
($km$)&($10^{11}\;Pa$)&($10^3\;kg\cdot m^{-3}$)&($K$)&$\left(10^4\;Pa
/m\right)$\\
\hline
1,216&3.300&12.20&4,961&-5.600\\
1,371&3.223&12.14&4,905&-6.094\\
1,571&3.094&12.03&4,824&-6.737\\
1,821&2.916&11.84&4,710&-7.507\\
2,171&2.636&11.52&4,521&-8.479\\
2,571&2.278&11.11&4,258&-9.421\\
2,971&1.886&10.62&3,936&-10.12\\
3,171&1.681&10.33&3,751&-10.34\\
3,371&1.473&10.01&3,551&-10.47\\
3,486&1.350&9.860&3,429&-10.56\\
\hline
\noalign{\hrule height 1.5pt}
\end{tabular}
\end{table}

   Since the activation volume increases strongly with pressure
(Brazhkin, 1998) as represented by our parameter $D$ in equation (\ref{eqn36}),
we allow for a linear variation with depth through
\bequ
D=D_b +\frac{b-r}{b-a}D_a,
\nequ
where $a$ is the radius of the inner core and $D_b$, $D_a$ are constants. The
integration in (\ref{eqn37}) is carried out by Simpson's rule over 100 steps
with spline interpolation across the whole outer core. It is found that the
constant, $D_b$, controls the curvature of the viscosity profile near the
core-mantle boundary, while the curvature otherwise departs only slightly from
log-linear. Some numerical experimentation shows that the profile is closely
log-linear, even near the core-mantle boundary, for a value $D_b=4.5\times
10^{-4}\;m^{-2}\cdot s^2\cdot K$ and that for $D_a=2.976\times 10^{-3}\;m^{-2}
\cdot s^2\cdot K$, the viscosity at the bottom of the outer core, $\eta_a=1.247
\times 10^{11}\;Pa\cdot s$, is closely matched. The resulting viscosity profile
is shown in Figure 8.
\begin{center}
\begin{figure}[h]
\hspace*{2.0cm}
\epsfbox{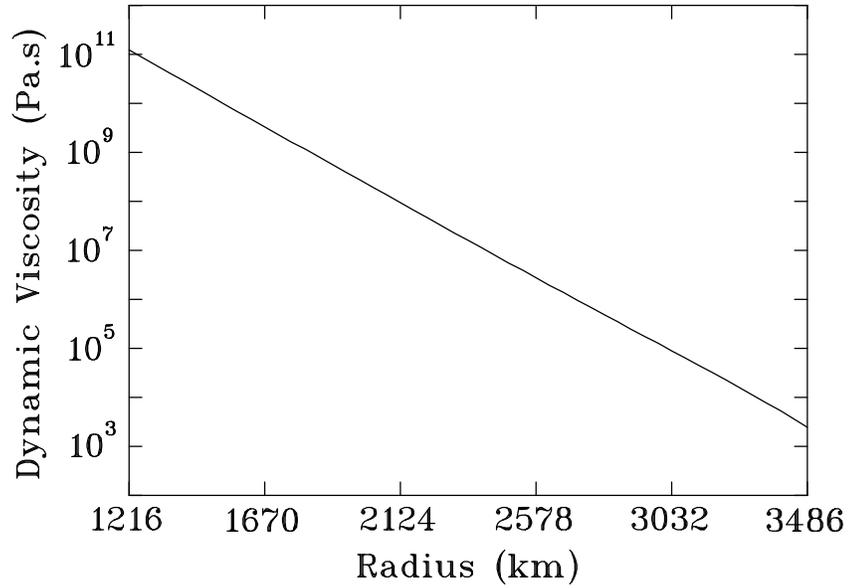}
\caption{Viscosity profile for Earth's Outer Core.}
\end{figure}
\end{center}
\newpage
   
\section{Discussion}

   Given past uncertainties, the agreement between viscosities in the outer
core measured from the VLBI observations of nutations, the superconducting
gravimeter observations of the translational modes, and the Arrhenius
extrapolation of laboratory high pressure and temperature experiments, is
quite remarkable. The very large gap between direct observations and
extrapolations of laboratory values appears to have closed. The viscosity
values involved are for molecular viscosities and they are large enough that
flows in the core are likely laminar in contradiction to conventional thought
that the flows are turbulent and that the large viscosities reflect eddy
viscosities.

   According to the viscosity profile we have derived, the local Ekman number
ranges from $1.2\times 10^{-2}$ at the bottom to $2.7\times 10^{-10}$ at the
top of the outer core. Our results appear to confirm the suggestion by
Braginsky (Braginsky, 1963) that the release of the latent heat of fusion as
metallic constituents freeze out in the F-layer at the bottom of the outer core
may be the energy source required to drive the geodynamo through compositional
convection as studied by Loper and Roberts (Loper and Roberts, 1981). At the
same time, except in the very lower part of the outer core, Ekman numbers are
in the range $10^{-4}$ to $10^{-5}$ ,or below, in which numerical dynamos
operate (Olsen and Christiansen, 2002).

\bigskip

\noindent
\large{\bfseries Acknowledgments}\\
\normalsize
\noindent
D.E.S. is grateful for financial support from
the Natural Sciences and Engineering Research Council of Canada. We are
indebted to Keith Aldridge for bringing the intricacies of Ekman layer
theory to our attention.

\newpage
\begin{center}
\Large{\bfseries{Appendix A}}
\end{center}
\large{\bfseries{Ekman Boundary Layers and Dissipation}}
\normalsize

   In this Appendix, we describe in detail the solution of the boundary layer
equations (\ref{eqn6}) and the calculation of the rates of energy dissipation
in the two boundary layers at the boundaries of the outer core. For convenience
in keeping track of phase, the velocity components
$\left(v_\theta,v_\phi\right)$ will be taken to be complex phasors with time
and longitude variations given by $e^{-i\left(\phi +\Omega t\right)}$. For an
assumed radial dependence proportional to $e^{\lambda r}$, we are lead to the
homogeneous system of equations
\bequ\label{eqn1}
\left(\begin{array}{ll}
\lambda^2\nu +i\Omega&2\Omega\cos\theta\\
-2\Omega\cos\theta&\lambda^2\nu +i\Omega\end{array}\right)
\left(\begin{array}{l}v_\theta\\
v_\phi\end{array}\right)=0.
\nequ        
For this system to have a solution, $\lambda$ must satisfy
\bequ\label{eqn3}
\left(\lambda^2\nu +i\Omega\right)^2=-4\Omega^2\cos^2\theta.
\nequ
Thus,
\bequ\label{eqn2}
\lambda^2\nu +i\Omega=\pm i2\Omega\cos\theta.
\nequ
Substitution of relation (\ref{eqn2}) into the system of equations
(\ref{eqn1}), gives
\bequ
v_\phi=\mp iv_\theta,
\nequ
and
\bequ
v_\phi=\pm iv_\phi.
\nequ

The boundary layers are characterized by the dimensionless Ekman number
\bequ
E_k=\frac{\nu}{b^2\Omega},
\nequ
with length scale fixed by the radius $b$ of the CMB. The four roots of the
secular equation (\ref{eqn3}) are $\pm\lambda_1,\;\pm\lambda_2$ with
\begin{align}
\lambda_1=&\frac{1-i}{b}\sqrt{\frac{1/2 +\cos\theta}{E_k}}=\frac{1+i}
{\delta_1}\;\;\;\mbox{for}\;\;\;\theta < 2\pi/3\label{eqn11}\\
=&\frac{1+i}{b}\sqrt{\frac{-1/2 -\cos\theta}{E_k}}=\frac{1+i}
{\delta_1}\;\;\;\mbox{for}\;\;\;\theta > 2\pi/3\label{eqn12}
\end{align}
and
\begin{align}
\lambda_2=&\frac{1-i}{b}\sqrt{\frac{1/2 -\cos\theta}{E_k}}=\frac{1-i}
{\delta_2}\;\;\;\mbox{for}\;\;\;\theta > \pi/3\label{eqn13}\\
=&\frac{1+i}{b}\sqrt{\frac{-1/2 +\cos\theta}{E_k}}=\frac{1+i}
{\delta_2}\;\;\;\mbox{for}\;\;\;\theta < \pi/3,\label{eqn14}
\end{align}
with $\delta_1,\delta_2$, the respective boundary layer thicknesses, both of
$O\left(\sqrt{E_k}\right)$.

   In the boundary layer near the top of the core, the perturbing velocity
components $\left(v_\theta,v_\phi\right)$ vanish with decreasing radius and
increase with radius, so that at $r=b$, they are equal and opposite to the
interior nutation velocity components (\ref{eqn4}) to satisfy the no-slip
condition at the CMB. Thus, the admissible values of $\lambda$ satisfying
equation (\ref{eqn2}) are $\lambda_1,\lambda_2$. In the boundary layer near
the bottom of the outer core, the perturbing velocity components vanish with
increasing radius and increase with decreasing radius to satisfy the no-slip
condition at the ICB. There, the admissible values of $\lambda$ satisfying
equation (\ref{eqn2}) are $-\lambda_1,-\lambda_2$. The perturbing velocity
components are then given by the linear combinations
\begin{align}
v_\theta=&e^{-\Delta r/\delta_1}\left(fe^{i\left(\Delta r
/\delta_1\pm\phi\pm\Omega t\right)}\right)+e^{-\Delta r/\delta_2}
\left(ge^{i\left(\Delta r/\delta_2\pm\phi\pm\Omega t\right)}
\right)\label{eqn7}\\
v_\phi=&e^{-\Delta r/\delta_1}\left(ife^{i\left(\Delta r
/\delta_1\pm\phi\pm\Omega t\right)}\right)-e^{-\Delta r/\delta_2}
\left(ige^{i\left(\Delta r/\delta_2\pm\phi\pm\Omega t\right)}
\right),\label{eqn8}
\end{align}
where the lower signs apply for the range $\pi/3<\theta<2\pi/3$, while the
upper signs apply for the range $0<\theta<\pi/3$ for terms involving
$\delta_2$, and the upper signs apply for the range $2\pi/3<\theta<\pi$ for
terms involving $\delta_1$. $\Delta r$ is the increment in radius. At the top
of the core $\Delta r=b-r$, and at the bottom of the outer core $\Delta r=r-a$.
In general, the linear combination coefficients are complex with real and
imaginary parts expressed by
\begin{align}
f=&\alpha +i\beta\\
g=&\gamma +i\epsilon.
\end{align}   

   Both the real and imaginary parts of expressions (\ref{eqn7}) and
(\ref{eqn8}) are solutions of the boundary layer equations (\ref{eqn6}).
Our interest is in the real parts of the velocity components given by
\begin{align}
Rl v_\theta=&e^{-\Delta r/\delta_1}\left[\left(\alpha\cos
\frac{\Delta r}{\delta_1}\pm\beta\sin\frac{\Delta r}{\delta_1}\right)\cos\left(
\phi +\Omega t\right)\right.\nonumber\\
+&\left.\left(\beta\cos\frac{\Delta r}{\delta_1}\mp\alpha\sin\frac{\Delta r}
{\delta_1}\right)\sin\left(\phi +\Omega t\right)\right]\nonumber\\
+&e^{-\Delta r/\delta_2}\left[\left(\gamma\cos\frac{\Delta r}
{\delta_2}\pm\epsilon\sin\frac{\Delta r}{\delta_2}\right)\cos\left(\phi +
\Omega t\right)\right.\nonumber\\
+&\left.\left(\epsilon\cos\frac{\Delta r}{\delta_2}\mp\gamma\sin\frac{\Delta r}
{\delta_2}\right)\sin\left(\phi +\Omega t\right)\right],\label{eqn9}
\end{align}
\begin{align}
Rl v_\phi=&e^{-\Delta r/\delta_1}\left[\left(-\beta\cos\frac{\Delta r}
{\delta_1}\pm\alpha\sin\frac{\Delta r}{\delta_1}\right)\cos\left(\phi +\Omega t
\right)\right.\nonumber\\
+&\left.\left(\alpha\cos\frac{\Delta r}{\delta_1}\pm\beta\sin\frac{\Delta r}
{\delta_1}\right)\sin\left(\phi +\Omega t\right)\right]\nonumber\\
+&e^{-\Delta r/\delta_2}\left[\left(\epsilon\cos\frac{\Delta r}
{\delta_2}\mp\gamma\sin\frac{\Delta r}{\delta_2}\right)\cos\left(\phi +\Omega t
\right)\right.\nonumber\\
+&\left.\left(-\gamma\cos\frac{\Delta r}{\delta_2}\mp\epsilon\sin
\frac{\Delta r}{\delta_2}\right)\sin\left(\phi +\Omega t\right)\right].
\label{eqn10}
\end{align}
For the velocity components (\ref{eqn9}) and (\ref{eqn10}) to cancel the
components of (\ref{eqn4}) at the boundaries, we have
\begin{align}
Rl v_\theta=&\left(\alpha +\gamma\right)\cos\left(\phi +\Omega t\right)+\left(
\beta +\epsilon\right)\sin\left(\phi +\Omega t\right)=Ar_0\sin\left(\phi +
\Omega t\right),\\
Rl v_\phi=&\left(\epsilon -\beta\right)\cos\left(\phi +\Omega t\right)+\left(
\alpha-\gamma\right)\sin\left(\phi +\Omega t\right)=Ar_0\cos\theta\cos\left(
\phi +\Omega t\right),
\end{align}
where $r_0$ is the boundary radius. $r_0=b$ at the top of the core and $r_0=a$
at the bottom of the outer core.
Then,
\bequ\label{eqn15}
\beta=\frac{1}{2}Ar_0\left(1-\cos\theta\right),\;\epsilon=\frac{1}{2}Ar_0\left(
1+\cos\theta\right),\;\alpha=\gamma=0.
\nequ
Differentiation of expressions (\ref{eqn9}) and (\ref{eqn10}) yields the
derivatives of the velocity components at the top of the core,
\begin{align}
\frac{\partial Rlv_\theta}{\partial r}=&\left(\mp\frac{\beta}{\delta_1}\mp
\frac{\epsilon}{\delta_2}\right)\cos\left(\phi +\Omega t\right)+\left(\frac{
\beta}{\delta_1}+\frac{\epsilon}{\delta_2}\right)\sin\left(\phi +\Omega t
\right),\\
\frac{\partial Rlv_\phi}{\partial r}=&\left(-\frac{\beta}{\delta_1}+\frac{
\epsilon}{\delta_2}\right)\cos\left(\phi +\Omega t\right)+\left(\mp\frac{\beta}
{\delta_1}\pm\frac{\epsilon}{\delta_2}\right)\sin\left(\phi +\Omega t\right),
\end{align}
while those at the bottom of the outer core are the negatives of these. At the
top of the core, the leading order stresses on the outer surface are
\bequ\label{eqn21}
\sigma_{r\theta}=\eta\frac{\partial Rlv_\theta}{\partial r}\;\;\;\mbox{and}
\;\;\;\sigma_{r\phi}=\eta\frac{\partial Rlv_\phi}{\partial r},
\nequ
where $\eta$ is the dynamic viscosity. At the bottom of the outer core, the
leading order stresses on the inner surface are the negatives of these, so that
on both surfaces the rate of dissipation of energy per unit area in the motion
with velocity (\ref{eqn4}) against these stresses is
\begin{align}
\frac{de}{dr}=&v_\theta\sigma_{r\theta}+v_\phi\sigma_{r\phi}\nonumber\\
=&Ar_0\eta\left[\left(\mp\frac{\beta}{\delta_1}\mp\frac{\epsilon}{\delta_2}
\right)\cos\left(\phi +\Omega t\right)\sin\left(\phi +\Omega t\right)+\left(
\frac{\beta}{\delta_1}+\frac{\epsilon}{\delta_2}\right)\sin^2\left(\phi +
\Omega t\right)\right.\nonumber\\
+&\left.\cos\theta\left(\left(\frac{\epsilon}{\delta_2}-\frac{\beta}{\delta_1}
\right)\cos^2\left(\phi +\Omega t\right)+\left(\mp\frac{\beta}{\delta_1}\pm
\frac{\epsilon}{\delta_2}\right)\cos\left(\phi +\Omega t\right)\sin\left(\phi
 +\Omega t\right)\right)
\right].
\end{align}
Substitution from (\ref{eqn15}), (\ref{eqn11}), (\ref{eqn12}), (\ref{eqn13})
and (\ref{eqn14}) yields, for $0<\theta<\pi/3$,
\begin{align}
\frac{de}{dt}=&\frac{1}{2}\rho_0 A^2 r_0^2\left[\sqrt{\nu\Omega\left(1/2+\cos
\theta\right)}\right.\nonumber\\
&\left.\left\{\sin^2\theta\left(\cos\left(\phi +\Omega t\right)\sin\left(\phi +
\Omega t\right)-\cos^2\left(\phi +\Omega t\right)\right)+1-\cos\theta\right\}
\right.\nonumber\\
&\left.-\sqrt{\nu\Omega\left(-1/2+\cos\theta\right)}\right.\nonumber\\
&\left.\left\{\sin^2\theta\left(\cos\left(\phi +\Omega t\right)\sin\left(\phi +
\Omega t\right)+\cos^2\left(\phi+\Omega t\right)\right)-1-\cos\theta\right\}
\right],
\end{align}
and for $\pi/3<\theta<2\pi/3$,
\begin{align}
\frac{de}{dt}=&\frac{1}{2}\rho_0 A^2 r_0^2\left[\sqrt{\nu\Omega\left(1/2 +\cos
\theta\right)}\right.\nonumber\\
&\left.\left\{\sin^2\theta\left(\cos\left(\phi +\Omega t\right)\sin\left(\phi +
\Omega t\right)-\cos^2\left(\phi +\Omega t\right)\right)+1-\cos\theta\right\}
\right.\nonumber\\
&\left.+\sqrt{\nu\Omega\left(1/2-\cos\theta\right)}\right.\nonumber\\
&\left.\left\{\sin^2\theta\left(\cos\left(\phi +\Omega t\right)\sin\left(\phi +
\Omega t\right)-\cos^2\left(\phi +\Omega t\right)\right)+1+\cos\theta\right\}
\right],
\end{align}
and for $2\pi/3<\theta<\pi$,
\begin{align}
\frac{de}{dt}=&\frac{1}{2}\rho_0 A^2r_0^2\left[-\sqrt{\nu\Omega\left(-1/2-\cos
\theta\right)}\right.\nonumber\\
&\left.\left\{\sin^2\theta\left(\cos\left(\phi +\Omega t\right)\sin\left(\phi +
\Omega t\right)+\cos^2\left(\phi +\Omega t\right)\right)-1+\cos\theta\right\}
\right.\nonumber\\
&\left.+\sqrt{\nu\Omega\left(1/2-\cos\theta\right)}\right.\nonumber\\
&\left.\left\{\sin^2\theta\left(\cos\left(\phi +\Omega t\right)\sin\left(\phi +
\Omega t\right)-\cos^2\left(\phi +\Omega t\right)\right)+1+\cos\theta\right\}
\right],
\end{align}
with $\rho_0$ representing the density just inside the respective boundary.

   Integrating over the entire boundary, the total rate of energy dissipation
is
\begin{align}
\frac{dE}{dt}=&r_0^2\int_0^{2\pi}\int_0^\pi\frac{de}{dt}\sin\theta d\theta d
\phi\nonumber\\
=&-\frac{\pi}{2}\rho_0 A^2 r_0^4\sqrt{\nu\Omega}\nonumber\\
&\times\left[\int_0^{\pi/3}\sqrt{\frac{1}{2}+\cos\theta}\left(\sin\theta\left(
1-\cos\theta +\sin^2\theta\right)\right)d\theta\right.\nonumber\\
&\left.+\int_0^{\pi/3}\sqrt{-\frac{1}{2}+\cos\theta}\left(\sin\theta\left(1+
\cos\theta+\sin^2\theta\right)\right)d\theta\right.\nonumber\\
&\left.+\int_{\pi/3}^{2\pi/3}\sqrt{\frac{1}{2}+\cos\theta}\left(\sin\theta
\left(1-\cos\theta +\sin^2\theta\right)\right)d\theta\right.\nonumber\\
&\left.+\int_{\pi/3}^{2\pi/3}\sqrt{\frac{1}{2}-\cos\theta}\left(\sin\theta
\left(1+\cos\theta +\sin^2\theta\right)\right)d\theta\right.\nonumber\\
&\left.+\int_{2\pi/3}^\pi\sqrt{-\frac{1}{2}-\cos\theta}\left(\sin\theta\left(1-
\cos\theta +\sin^2\theta\right)\right)d\theta\right.\nonumber\\
&\left.+\int_{2\pi/3}^\pi\sqrt{\frac{1}{2}-\cos\theta}\left(\sin\theta\left(1+
\cos\theta +\sin^2\theta\right)\right)d\theta\right],\label{eqn16}
\end{align}
where, in the integration over $\phi$, we have made use of the integrals
\bequ
\int_0^{2\pi}\cos\left(\phi +\Omega t\right)\sin\left(\phi +\Omega t\right)d
\phi=\frac{1}{2}\int_0^{2\pi}\sin\left\{2\left(\phi+\Omega t\right)\right\}d
\phi=0,
\nequ
and
\bequ
\int_0^{2\pi}\cos^2\left(\phi +\Omega t\right)d\phi=\frac{1}{2}\int_0^{2\pi}
\left(\cos\left\{2\left(\phi +\Omega t\right)\right\}+1\right)d\phi=\pi.
\nequ
We may write (\ref{eqn16}) in the shorthand
\bequ\label{eqn23}
\frac{dE}{dt}=-\frac{\pi}{2}\rho_0 A^2 r_0^4\sqrt{\nu\Omega}\left[J_1 +J_2 +J_3
+J_4 +J_5 +J_6\right],
\nequ
with $J_1,J_2,J_3,J_4,J_5,J_6$ representing the six integrals in (\ref{eqn16}).
Evaluation of the integrals in expression (\ref{eqn16}) depends on the
indefinite integrals
\begin{align}
&I_1=\int\left(\frac{s}{2}\mp\cos\theta\right)^{1/2}\sin^3\theta 
d\theta\nonumber\\
=&\pm\frac{2}{3}\left(\frac{s}{2}\mp\cos\theta\right)^{3/2}\sin^2
\theta-\frac{8}{15}\left(\frac{s}{2}\mp\cos\theta\right)^{5/2}\cos
\theta\mp\frac{16}{105}\left(\frac{s}{2}\mp\cos\theta\right)^{7/2},
\label{eqn17}\\
&I_2=\int\left(\frac{s}{2}\mp\cos\theta\right)^{1/2}\left(1-\cos\theta
\right)\sin\theta d\theta\nonumber\\
=&\pm\frac{2}{3}\left(\frac{s}{2}\mp\cos\theta\right)^{3/2}\mp\frac{2}{3}\left(
\frac{s}{2}\mp\cos\theta\right)^{3/2}\cos\theta -\frac{4}{15}\left(\frac{s}{2}
\mp\cos\theta\right)^{5/2},\label{eqn18}\\
&I_3=\int\left(\frac{s}{2}\mp\cos\theta\right)^{1/2}\left(1+\cos\theta
\right)\sin\theta d\theta\nonumber\\
=&\pm\frac{2}{3}\left(\frac{s}{2}\mp\cos\theta\right)^{3/2}\pm\frac{2}{3}\left(
\frac{s}{2}\mp\cos\theta\right)^{3/2}\cos\theta +\frac{4}{15}\left(\frac{s}{2}
\mp\cos\theta\right)^{5/2},\label{eqn19}
\end{align}
$s$ representing the sign, which can take on either the value $+1$, or the
value $-1$, throughout each expression. Using (\ref{eqn17}), (\ref{eqn18}) and
(\ref{eqn19}), and inserting limits of integration, we find
\bequ
J_1=\frac{13}{30}+\frac{16}{105}-\frac{18}{35}\sqrt{\frac{3}{2}},\;\;J_2=-
\frac{57}{105}\sqrt{2},\;\;J_3=-\frac{1}{6}-\frac{44}{105},
\nequ
\bequ
J_4=-\frac{1}{6}-\frac{44}{105}=J_3,\;\;J_5=-\frac{57}{105}\sqrt{2}=J_2,\;\;J_6
=\frac{13}{30}+\frac{16}{105}-\frac{18}{35}\sqrt{\frac{3}{2}}=J_1.
\nequ
The six integrals sum to
\bequ
J_1 +J_2 +J_3 +J_4 +J_5 +J_6=-2\sqrt{2}\left(9\sqrt{3}+19\right)/35.
\nequ
Substituting this result in expression (\ref{eqn23}), we find the rate of
energy dissipation in each of the respective boundary layers to be
\bequ\label{eqn24}
\frac{dE}{dt}=\frac{\pi}{35}\rho_0 A^2 r_0^4\sqrt{2\nu\Omega}\left(9\sqrt{3}+19
\right).
\nequ
\newpage
\begin{center}
\Large{\bfseries{Appendix B}}
\end{center}

\large{\bfseries{Viscous Coupling to the Inner Core and Shell}}
\normalsize

   From relation (\ref{eqn20}), it is apparent that the reciprocal of the
overall quality factor cannot be less than the reciprocal of the effective
quality factor arising from the boundary layer at the ICB. Because of the
high viscosity there, the inner core is likely to be tightly coupled to
outer core wobble. The nearly retrograde diurnal wobbles of the outer core,
associated with the free core nutations, give rise to viscous torques exerted
by the outer core on the inner core and shell.

   The viscous torques exerted by the outer core at the boundaries are
\bequ
{\bm \Gamma}=\int {\bm r}\times\left(\hat{\bm \theta}\sigma_{r\theta}+\hat{
\bm \phi}\sigma_{r\phi}\right)dS=\int\left(-\hat{\bm \theta}r\sigma_{r\phi}
+\hat{\bm \phi}r\sigma_{r\theta}\right)dS,
\nequ
where the integral is over the respective boundary surface. The spherical
polar unit vectors are related to the Cartesian unit vectors $\left(\hat{
\bm i},\hat{\bm j},\hat{\bm k}\right)$ by
\bequ
\hat{\bm \theta}=\hat{\bm i}\cos\theta\cos\phi+\hat{\bm j}\cos\theta\sin\phi-
\hat{\bm k}\sin\theta,
\nequ
\bequ
\hat{\bm \phi}=-\hat{\bm i}\sin\phi+\hat{\bm j}\cos\phi.
\nequ 
The Cartesian components of the viscous torques are then
\begin{align}
{\bm \Gamma}=&\hat{\bm i}r_0^3\int_0^{2\pi}\int_0^\pi\sin\theta\left(-\sigma_
{r\phi}\cos\theta\cos\phi -\sigma_{r\theta}\sin\phi\right)d\theta d\phi
\nonumber\\
+&\hat{\bm j}r_0^3\int_0^{2\pi}\int_0^\pi\sin\theta\left(-\sigma_{r\phi}\cos
\theta\sin\phi+\sigma_{r\theta}\cos\phi\right)d\theta d\phi\nonumber\\
+&\hat{\bm k}r_0^3\int_0^{2\pi}\int_0^\pi\sigma_{r\phi}\sin^2\theta d\theta d
\phi.
\end{align}
From the expressions (\ref{eqn21}) for the leading order stresses, the
integrations over $\phi$ are seen to depend on the elementary integrals
\begin{align}
&\int _0^{2\pi}\cos\left(\phi +\Omega t\right)\cos\phi d\phi=\pi\cos\Omega t,
\;\;\;\int_0^{2\pi}\sin\left(\phi +\Omega t\right)\cos\phi d\phi=\pi\sin\Omega
t,\nonumber\\
&\int_0^{2\pi}\cos\left(\phi +\Omega t\right)\sin\phi d\phi=-\pi\sin\Omega t,
\;\;\;\int_0^{2\pi}\sin\left(\phi +\Omega t\right)\sin\phi d\phi=\pi\cos\Omega
t,\nonumber\\
&\int_0^{2\pi}\cos\left(\phi +\Omega t\right)d\phi=\int_0^{2\pi}\sin\left(\phi
+\Omega t\right)d\phi=0.\nonumber
\end{align}  
The latter two integrals ensure that the viscous torques have only equatorial
Cartesian components, $\left(\Gamma_x,\Gamma_y\right)$, and writing $\tilde{
\Gamma}=\Gamma_x +i\Gamma_y$, we have
\begin{align}
\tilde{\Gamma}=&-\pi r_0^3\eta e^{-i\Omega t}\int_0^\pi\left[\cos\theta\sin
\theta\left(\frac{\beta}{\delta_1}-\frac{\epsilon}{\delta_2}\right)-\sin\theta
\left(\frac{\beta}{\delta_1}+\frac{\epsilon}{\delta_2}\right)\right]d\theta
\nonumber\\
+&i\pi r_0^3\eta e^{-i\Omega t}\int_0^\pi\left[\cos\theta\sin\theta\left(\mp
\frac{\beta}{\delta_1}\pm\frac{\epsilon}{\delta_2}\right)+\sin\theta\left(\pm
\frac{\beta}{\delta_1}\pm\frac{\epsilon}{\delta_2}\right)\right]d\theta.
\label{eqn22}
\end{align}
Substitution from equations (\ref{eqn11}), (\ref{eqn12}), (\ref{eqn13}),
(\ref{eqn14}) and (\ref{eqn15}) shows that evaluation of the torque expression
(\ref{eqn22}) depends on the integrals
\begin{align}
&\int\sqrt{\left(\frac{s}{2}\mp\cos\theta\right)}\cos\theta\sin\theta d\theta
\nonumber\\
&=\pm\frac{2}{3}\cos\theta\left(\frac{s}{2}\mp\cos\theta\right)^{3/2}+\frac{
4}{15}\left(\frac{s}{2}\mp\cos\theta\right)^{5/2},\nonumber\\
&\int\sqrt{\left(\frac{s}{2}\mp\cos\theta\right)}\cos^2\sin\theta d\theta
\nonumber\\
&=\pm\frac{2}{3}\left(\frac{s}{2}\mp\cos\theta\right)^{3/2}\cos^2\theta +
\frac{8}{15}\left(\frac{s}{2}\mp\cos\theta\right)^{5/2}\cos\theta\pm\frac{16}
{105}\left(\frac{s}{2}\mp\cos\theta\right)^{7/2},\nonumber\\
&\int\sqrt{\left(\frac{s}{2}\mp\cos\theta\right)}\sin\theta d\theta=
\pm\frac{2}{3}\left(\frac{s}{2}\mp\cos\theta\right)^{3/2},\nonumber\\
\end{align}
where, again, $s$ represents the sign, which can take on either the value $+1$,
or the value $-1$, throughout each expression.

The total viscous torque is made up of contributions from three zones of
latitude. The contribution from the region $0<\theta<\pi/3$ is
\begin{align}
\tilde{\Gamma}_{0<\theta<\pi/3}=&-\pi\rho_0 Ar_0^4\sqrt{\nu
\Omega}e^{-i\Omega t}\left[\frac{41}{140}-\frac{9}{35}\sqrt{\frac{3}{2}}-\frac{
19}{35}\frac{1}{\sqrt{2}}\right.\nonumber\\
&\left.+i\left\{-\frac{41}{140}+\frac{9}{35}\sqrt{\frac{3}{2}}
-\frac{19}{35}\frac{1}{\sqrt{2}}\right\}\right],
\end{align}
while the contribution from the region $\pi/3<\theta<2\pi/3$ is
\bequ
\tilde{\Gamma}_{\pi/3<\theta<2\pi/3}=-\pi\rho_0 Ar_0^4\sqrt{\nu
\Omega}e^{-i\Omega t}\frac{41}{70}\left(-1+i\right),
\nequ
and that from the region $2\pi/3<\theta<\pi$ is
\begin{align}
\tilde{\Gamma}_{2\pi<\theta<\pi}=&-\pi\rho_0 Ar_0^4\sqrt{\nu
\Omega}e^{-i\Omega t}\left[\frac{41}{140}-\frac{9}{35}\sqrt{\frac{3}{2}}-\frac{
19}{35}\frac{1}{\sqrt{2}}\right.\nonumber\\
&\left.+i\left\{-\frac{41}{140}+\frac{9}{35}\sqrt{\frac{3}{2}}-\frac{19}{35}
\frac{1}{\sqrt{2}}\right\}\right].
\end{align}
The total viscous torque is then
\bequ\label{eqn25}
\tilde{\Gamma}=\pi\rho_0 Ar_0^4\sqrt{\nu\Omega}e^{-i\Omega t}
\frac{19}{35}\sqrt{2}\left[1+i+\frac{9}{19}\sqrt{3}\left(1-i\right)
\right].
\nequ
Separating the torques from the two boundary layers, the outer core exerts the
viscous torque
\bequ\label{eqn26}
\tilde{\Gamma}_a=\pi\rho_0\left(a\right)A_a a^4\sqrt{\nu_a \Omega}e^
{-i\Omega t}\frac{19}{35}\sqrt{2}\left[1+i+\frac{9}{19}\sqrt{3}\left(1-i\right)
\right]
\nequ
on the inner core, while it exerts the viscous torque
\bequ\label{eqn27}
\tilde{\Gamma}_b=\pi\rho_0\left(b\right)A_b b^4\sqrt{\nu_b\Omega}e^
{-i\Omega t}\frac{19}{35}\sqrt{2}\left[1+i+\frac{9}{19}\sqrt{3}\left(1-i\right)
\right]
\nequ
on the shell.

   The extra nearly diurnal retrograde wobble of the outer core compared to its
boundaries, in complex phasor notation, is
\bequ
\tilde{\omega}=\omega_1 +i\omega_2=Ae^{-i\Omega t},
\nequ
with wobble angular velocity ${\bm \omega}=\left(\omega_1,\omega_2\right)$.
The rate at which the outer core does work against the viscous torques is
\bequ
\frac{dE}{dt}={\bm \Gamma}\cdot{\bm \omega}=\Gamma_1\omega_1 +\Gamma_2\omega_2
=\frac{1}{2}\left(\tilde{\omega}\tilde{\Gamma}^\ast +\tilde{\omega}^\ast\tilde
{\Gamma}\right),
\nequ
where the torque vector is ${\bm \Gamma}=\left(\Gamma_1,\Gamma_2\right)$, and
where the superscript asterisk indicates the complex conjugate. From the torque
expression (\ref{eqn25}), we find the rate of energy dissipation in the
boundary layers to be
\bequ
\frac{dE}{dt}=\frac{\pi}{35}\rho_0 A^2 r_0^4\sqrt{2\nu\Omega}\left(9\sqrt{3}+
19\right),\nequ
in agreement with equation (\ref{eqn24}).

    The expressions (\ref{eqn26}) and (\ref{eqn27}) for the viscous torques
the outer core exerts on its boundaries may be abbreviated to
\bequ
\tilde{\Gamma}_a=\gamma_a e^{-i\Omega t}A_a
\nequ
and
\bequ
\tilde{\Gamma}_b=\gamma_b e^{-i\Omega t}A_b,
\nequ
where
\bequ\label{eqn30}
\gamma_a=\pi\rho_0\left(a\right)a^4\sqrt{\nu_a\Omega}\frac{19}{35}\sqrt{2}
\left[1+i+\frac{9}{19}\sqrt{3}\left(1-i\right)\right]
\nequ
and
\bequ\label{eqn31}
\gamma_b=\pi\rho_0\left(b\right)b^4\sqrt{\nu_b\Omega}\frac{19}{35}\sqrt{2}
\left[1+i+\frac{9}{19}\sqrt{3}\left(1-i\right)\right].
\nequ

\end{document}